\documentclass[sigconf]{acmart}
\usepackage[utf8]{inputenc} % allow utf-8 input
\usepackage[T1]{fontenc}    % use 8-bit T1 fonts
\usepackage{booktabs}       % professional-quality tables
\usepackage{amsfonts}       % blackboard math symbols
\usepackage{amsmath}        % math stuff
\usepackage{nicefrac}       % compact symbols for 1/2, etc.
\usepackage{microtype}      % microtypography
\usepackage{lipsum}		    % Can be removed after putting your text content
\usepackage{graphicx}
\usepackage{xcolor}         % coloring
\usepackage{setspace}       % spacing
\usepackage{subcaption}     % multiple plots in one row
\usepackage{verbatim}       % multiline comments
\usepackage{placeins}       % for getting floatbarriers
\usepackage{hyperref}
\usepackage{algorithm} 
\usepackage{listings}
\usepackage{algpseudocode}

\usepackage{color}

\makeatletter
\renewcommand{\Function}[2]{%
  \csname ALG@cmd@\ALG@L @Function\endcsname{#1}{#2}%
  \def\jayden@currentfunction{#1}%
}
\newcommand{\funclabel}[1]{%
  \@bsphack
  \protected@write\@auxout{}{%
    \string\newlabel{#1}{{\jayden@currentfunction}{\thepage}}%
  }%
  \@esphack
}

\AtBeginDocument{%
  \providecommand\BibTeX{{%
    \normalfont B\kern-0.5em{\scshape i\kern-0.25em b}\kern-0.8em\TeX}}}

\copyrightyear{2023}
\acmYear{2023}
\setcopyright{rightsretained}
\acmConference[ICAIF '23]{4th ACM International Conference on AI in Finance}{November 27--29, 2023}{Brooklyn, NY, USA}
\acmBooktitle{4th ACM International Conference on AI in Finance (ICAIF '23), November 27--29, 2023, Brooklyn, NY , USA} \acmDOI{10.1145/3604237.3626875} \acmISBN{979-8-4007-0240-2/23/11}

\begin{document}

%%
%% The "title" command has an optional parameter,
%% allowing the author to define a "short title" to be used in page headers.
%\title{Sparsification and Filtering for Spatial-temporal GNN in Multivariate Time-series}%%
\title{Topological Portfolio Selection and Optimization}
%% The "author" command and its associated commands are used to define
%% the authors and their affiliations.
%% Of note is the shared affiliation of the first two authors, and the
%% "authornote" and "authornotemark" commands
%% used to denote shared contribution to the research.
\author{Yuanrong Wang}
\affiliation{
  \institution{University College London}
  \streetaddress{66-72 Gower Street}
  \city{London}
  \country{UK}}
\email{yuanrong.wang@cs.ucl.ac.uk}

\author{Antonio Briola}
\affiliation{
  \institution{University College London}
  \streetaddress{66-72 Gower Street}
  \city{London}
  \country{UK}}
\email{antonio.briola.20@ucl.ac.uk}

\author{Tomaso Aste}
\authornote{Corresponding author.}
\affiliation{
  \institution{University College London}
  \streetaddress{66-72 Gower Street}
  \city{London}
  \country{UK}}
\email{t.aste@ucl.ac.uk}

%%
%% By default, the full list of authors will be used in the page
%% headers. Often, this list is too long, and will overlap
%% other information printed in the page headers. This command allows
%% the author to define a more concise list
%% of authors' names for this purpose.
\renewcommand{\shortauthors}{Y.Wang et.al.}

%%
%% The abstract is a short summary of the work to be presented in the
%% article.
\begin{abstract}
\noindent Modern portfolio optimization is centered around creating a low-risk portfolio with extensive asset diversification. Following the seminal work of Markowitz, optimal asset allocation can be computed using a constrained optimization model based on empirical covariance. However, covariance is typically estimated from historical lookback observations, and it is prone to noise and may inadequately represent future market behavior. As a remedy, information filtering networks from network science can be used to mitigate the noise in empirical covariance estimation, and therefore, can bring added value to the portfolio construction process. In this paper, we propose the use of the Statistically Robust Information Filtering Network (SR-IFN) which leverages the bootstrapping techniques to eliminate unnecessary edges during the network formation and enhances the network's noise reduction capability further. We apply SR-IFN to index component stock pools in the US, UK, and China to assess its effectiveness. The SR-IFN network is partially disconnected with isolated nodes representing lesser-correlated assets, facilitating the selection of peripheral, diversified and higher-performing portfolios. Further optimization of performance can be achieved by inversely proportioning asset weights to their centrality based on the resultant network.

%In the realm of contemporary quantitative finance, portfolio optimization holds paramount significance. The construction of a profitable, low-risk portfolio is primarily hinged on selecting highly diversified assets, a process typically reliant on the observation of the empirical covariance and correlation. The advent of network science has ushered in numerous network filtering methodologies that bolster the estimation of correlation and covariance matrices by mitigating noise interference within historical data. In this paper, we propose the Statistically Robust Information Filtering Network (SR-IFN) which leverages the bootstrapping techniques to eliminate unnecessary edges during the network formation process. Assets that remain disconnected within the resulting network constructed from index component stocks pool are utilized to establish an equally weighted peripheral portfolio, thereby effectuating considerable enhancement to the index, even when accounting for transaction costs. Further optimization of performance can be achieved by inversely proportioning asset weights to their centrality based on the resultant network. Experiments are performed over a 10-year period in the US, UK and China, which shows a statistically significant and robust boost in Sharpe Ratio and resilience to extreme market dynamics.
\end{abstract}

%%
%% The code below is generated by the tool at http://dl.acm.org/ccs.cfm.
%% Please copy and paste the code instead of the example below.
%%


%%
%% Keywords. The author(s) should pick words that accurately describe
%% the work being presented. Separate the keywords with commas.

%% A "teaser" image appears between the author and affiliation
%% information and the body of the document, and typically spans the
%% page.
%\begin{teaserfigure}
%  \includegraphics[width=\textwidth]{sampleteaser}
%  \caption{Seattle Mariners at Spring Training, 2010.}
%  \Description{Enjoying the baseball game from the third-base
%  seats. Ichiro Suzuki preparing to bat.}
%  \label{fig:teaser}

\maketitle
\section{Introduction}\label{sec:intro}

The optimization of financial portfolios has long been a focal point of investigation within the domains of finance and quantitative trading. The first mathematical formulation of the problem follows the seminal works of Markowitz in the 1950s \cite{Stuart1959PortfolioSE, Markowitz}. Following Markowitz, an optimal portfolio that minimizes variance for a given expected return is to be found by solving a quadratic optimization problem under linear constraint, and the closed-form solutions form the efficient frontier. The minimum variance portfolio (MVP) lies on the efficient frontier line minimizing the variance, and it is widely regarded and practiced by academia and industry as the most classic solution to the optimization problem. The MVP solution is simple and elegant that is contingent solely on the covariance of assets' historical return, independent of the mean. The covariance captures the volatility of a single asset and the interdependencies (correlation) between them. However, empirical covariance is notably hard to estimate, particularly within multivariate financial time series where the signal-to-noise ratio is exceptionally low. Consequently, minor perturbations can trigger significant deviations in portfolio construction. Additionally, financial markets are frequently subject to shifts and jumps, rendering historical empirical covariance an unsatisfactory predictor of future trends, and the prediction of future covariance a challenging task. These factors unfortunately invalidate the mathematical optimality of the MVP.

Recent progress in network science, especially in network filtering, has provided alternatives to the traditional covariance-based methodologies. The covariance and correlation matrices can be interpreted as a graph/network and can be condensed to essential information under certain graphical constraints, such as the minimum spanning tree (MST). The resulting filtered matrix typically exhibits sparsity with many structural zeroes, which correspond to statistically insignificant network components, thereby enhancing the matrix's robustness and generality. The resulting sparse network has been demonstrated to be useful for visualization and to aptly reflect market dynamics \cite{Mantegna1999HierarchicalSI, Onnela2003DynamicsOM}. Consequently, investment decisions, including portfolio selection and optimization, can be based on this network. For instance, the positive-defined sparse inverse covariance matrix from Triangulated Maximally Filtered Graph (TMFG) can directly substitute the original empirical inverse covariance matrix in the Markowitz model, yielding substantial improvements \cite{Wang2021DynamicPO, Procacci2021PortfolioOW}. Other topological information, such as centrality and peripherality, and community clusters, can be used as criteria for stock selection and portfolio weight optimization.

Information Filtering Networks (IFN) represents a robust and computationally efficient network filtering technique. However, due to certain topological constraints necessary for network construction, superfluous edges need to be introduced, which leads to slightly increasing the amount of noise. This paper presents a novel method, the Statistically Robust Information Filtering Network (SR-IFN), which employs a statistically robust bootstrapping approach to mitigate the noise introduced during the IFN's building pipeline. In this method, the underlying multivariate time series is bootstrapped multiple times and transformed into sparse sub-networks. These sub-networks are then ensembled, and only the key structures that occur more frequently than a predefined threshold are retained. This strategy prunes unnecessary components, increasing the informativeness of the remaining edges and maximizing the likelihood of the modeled system. This enhanced sparse network is then used for portfolio selection based on connectivity, as a peripheral portfolio is more diverse and entails lower risk. Further optimization can be carried out using the centrality of assets as a measure for weight calculation. We conduct experiments utilizing the component stocks of NASDAQ, FTSE, and HS300, representative of the equity markets in the US, UK, and China, respectively, and we include both scenarios with and without 20 basis point transaction costs.

The remainder of this paper is structured as follows. Section \ref{sec:litrev} presents a review of existing literature on network-based portfolio and information filtering networks. Then, a detailed explanation of SR-IFN is introduced in Section \ref{sec:SSifn}, and the associated methods for portfolio selection and optimization are included in Section \ref{sec:portSlcOpt}. Implementation details are shown in Section \ref{sec:imp}. Final results for selection and optimization are showcased and discussed in Section \ref{sec:portSelectResults} and Section \ref{sec:portOptResults}.

\section{Literature}\label{sec:litrev}
\subsection{Network-based Portfolio Optimization}\label{sec:lt_pso}
Traditional methods of portfolio optimization largely rely on empirical covariance and correlation, which predominantly capture linear dependencies among assets. Nevertheless, financial market networks synthesized from historical data, tend to encapsulate the entire system's complexity, including non-linearities, and often yield superior outcomes in terms of portfolio construction. Pozzi et al. \cite{Pozzi2013SpreadOR} in 2013 found that risk is not uniformly distributed across the market, with peripheral assets of a financial network demonstrating greater success in diversification and leading to superior performance. This finding has subsequently led research to focus on quantifying peripherality and constructing highly diversified, low-risk portfolios. To quantify peripherality, a graph must be initially treated using network filtering methodologies, such as information filtering networks, which transform the complete graph constructed from the correlation matrix or other linear \cite{Li2013UnveilingCB} and non-linear \cite{Sensoy2014DynamicST} similarity measures to a sparse network retaining only strongest relationships. Subsequently, different centrality measures, including degree centrality, betweenness centrality, eccentricity, and closeness centrality, are computed for each node. Nodes are then ranked in ascending order to be incorporated into the portfolio with equal or Markowitz weights \cite{Li2019PortfolioOB}, or weights that are calculated based on the centrality measures \cite{Peralta2016ANA}. Additional research includes network-based allocation with machine learning \cite{Konstantinov2020ANA}, cross-sectional equity sector portfolio construction \cite{Lohre2014TheUO}, and graph clustering-based portfolio construction \cite{Dees2019PortfolioCA}.

\subsection{Information Filtering Network (IFN)}

In recent years, substantial advancements have been observed in complex systems-driven data scrutiny through the utilization of information-filtering networks. This methodology depicts the interactions within intricate systems as network architectures composed of elements, or vertices, and interactions, or edges. A renowned technique initially proposed by Boruvka in 1926, the Minimum Spanning Tree (MST), can be accurately resolved via diverse methodologies \cite{nevsetvril2001otakar, CompNet10, CompNet11}. The MST condenses the architecture into a connected tree while preserving the more significant correlations.
With an intent to extract higher value data more efficaciously, both Tumminello et al. \cite{CompNet3} and Aste and Di Matteo \cite{CompNet4} suggested the incorporation of planar graphs within the Planar Maximally Filtered Graph (PMFG) algorithm. A planar graph can be embedded in the Euclidean plane so that no edges intersect. Subsequent research has extended this strategy to include chordal graphs, a type of graph in which all cycles of four or more vertices have a chord, which is an edge that is not part of the cycle but connects two vertices of the cycle, that exhibits variable sparsity \cite{CompNet5, CompNet6}. This approach has found applications in a variety of research fields, such as finance \cite{CompNet7} and neural systems \cite{CompNet8}, offering a robust mechanism for deciphering high-dimensional dependencies and building sparse representations. It has been empirically proven that chordal information filtering networks, inclusive of the Triangulated Maximally Filtered Graph (TMFG) \cite{CompNet5}, can yield a sparse precision matrix that is positive definite and encapsulates the network's structure, thereby enabling effective $L_0$-norm topological regularization \cite{aste2020topological}. Further explorations of the Maximally Filtered Clique Forest (MFCF) \cite{Massara2019LearningCF} have extended the technique's range of application by adapting it to cliques of various dimensions. This methodology has been demonstrated to be more computationally efficient and stable than Graphical LASSO \cite{CompNet9} and covariance shrinkage approaches \cite{Ledoit2003HoneyIS, Ledoit2004AWE, Briola2022DependencySI}, particularly in situations where data points are scarce \cite{CompNet7, CompNet4, Wang2021DynamicPO, Saef2022RegimebasedIS}.

\section{Methodologies}
\subsection{Statistically Robust IFN (SR-IFN)}\label{sec:SSifn}

The application of Information Filtering Networks (IFNs) has been extensively explored within the field of finance, particularly for the purpose of correlation/covariance sparsification and filtering. Nevertheless, given that IFNs specify a complete network/graph under certain topological constraints, e.g., planarity for PMFG, and chordality for TMFG, the resultant network structure incorporates elements that, while necessary to uphold these constraints, are irrelevant to the original information. This paper introduces a Statistically Robust (SR) method aimed at enhancing the stability and performance of IFNs by endeavoring to eliminate these constraint-related structures. Triangulated Maximally Filtered Graph (TMFG) is employed as the core IFN for the purposes of the ensuing experiments.

\begin{algorithm}
    \caption{Statistically Robust IFN (SR-IFN)}\label{alg:SSIFN}
    \textbf{Input} A set of observations $\mathbf{\hat{x}}_{s, n}  \in \mathbb{R}^{s, n}$, the confidence level $p_{cl}$, and the number of repetitions $t_{r}$\\
    \textbf{Output} Sparse similarity matrix $\mathbf{{S}}$.

    \vspace{0.05cm}
    
    \begin{algorithmic}[1]

        \State Initialize an empty ensemble adjacency matrix $\mathbf{A} \in \mathbb{R}^{n, n}$ with all zeros;

        \State Initialize an empty final sparse similarity matrix $\mathbf{{S}} \in \mathbb{R}^{n, n}$ with all zeros;

        \State Calculate the original correlation matrix $\hat{\mathbf{C}}$ $\in \mathbb{R}^{n, n}$ from $\mathbf{\hat{x}}_{s, n}$;

        \For{$t \gets 1$ to $t_r$} 
            \State Randomly bootstrap $\mathbf{\hat{x}}_{s, n}$ in the first dimension and obtain bootstrapped $\mathbf{\hat{x}}_{s, n}^t$;
            
             \State Calculate the bootstrapped correlation $\hat{\mathbf{C}}^t$ $\in \mathbb{R}^{n, n}$ from $\mathbf{\hat{x}}_{s, n}^t$;
        
            \State Obtain the bootstrapped sparse adjacency matrix $\mathbf{A}^t$ from $\hat{\mathbf{C}}^t$ by TMFG (or any information filtering network);

            \State $\mathbf{A} += \mathbf{A}^t$

        \EndFor
        \For {each pair of nodes $i, j$ in $\mathbf{A}$}
                \If {$\frac{A_{i,j}}{t_r} >  p_{cl}$}
                    \State $S_{i,j}$ = $\hat{{C}}_{i,j}$;
                \EndIf
        \EndFor

        \State \Return $\mathbf{{S}}$.
    \end{algorithmic}
\end{algorithm}

The construction process for TMFG relies on a simple topological move that maintains both planarity and chordality. TMFG has been demonstrated to be a computationally efficient model capable of sparse probabilistic modelling via topological regularization. However, it is not without limitations: unnecessary edges may be added to satisfy the graph's chordality, thereby introducing undesirable noise, a particular issue in fields characterized by a low signal-to-noise ratio, such as finance. To address this limitation, we propose the Statistically Robust (SR) method, detailed in Algorithm \ref{alg:SSIFN}.

\begin{figure}[h!]
    \centering
    \includegraphics[width=0.45\textwidth]{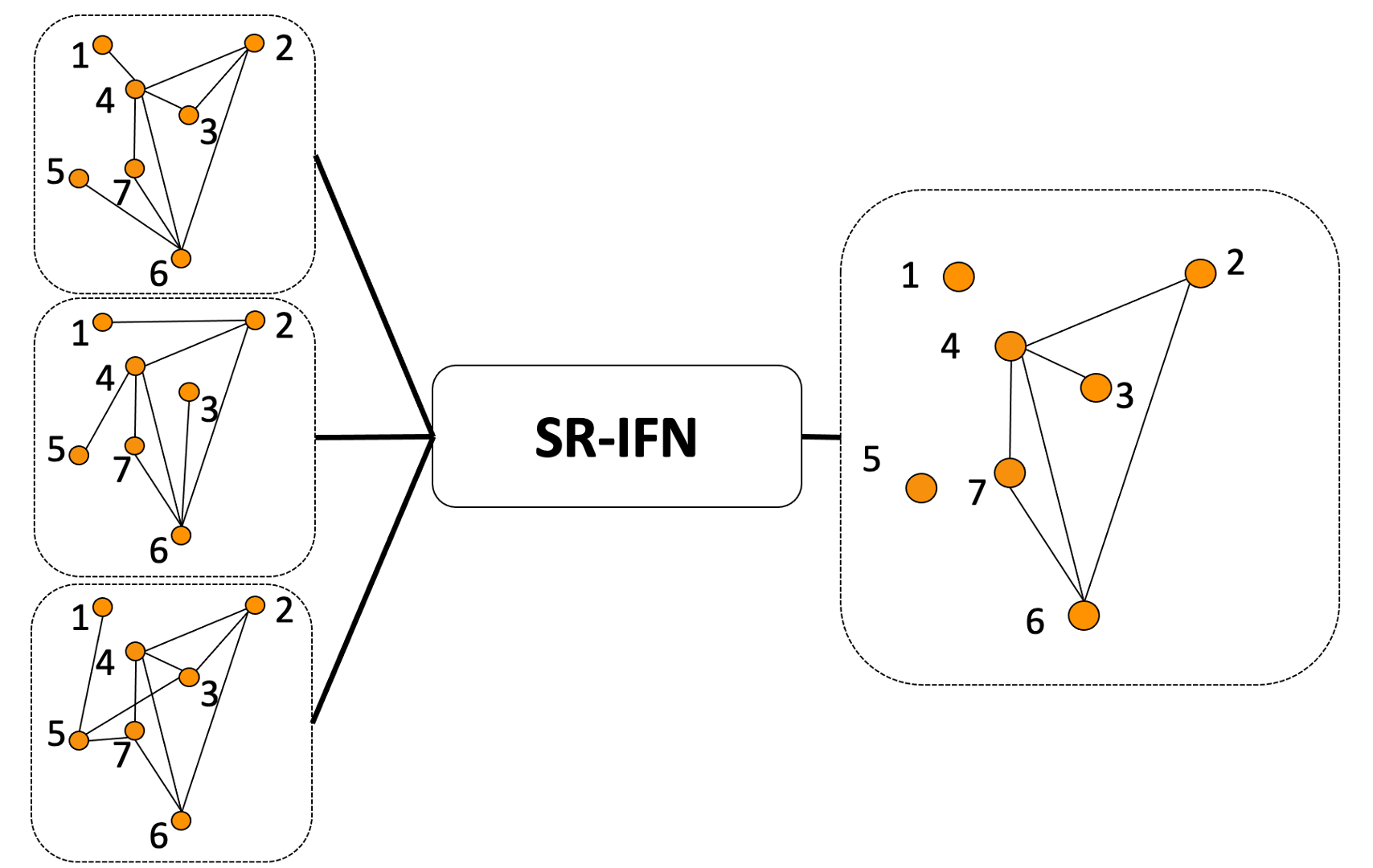}

    \caption{Statistically robust bootstrapping process. Three sub-networks were generated from one observation set with bootstrapping. Only edges that present more than a two-thirds majority will be preserved in the resulting statistically robust network.
}\label{fig:srbp}
\end{figure}

Temporal sequential dependence is reduced by randomly bootstrapping the observations in each repetition, which also results in each bootstrapped sample possessing a distinct network structure. Therefore, superfluous edges will be added differently in each case. By retaining structures that appear more frequently than a certain threshold, we can discard unnecessary edges and noise as they lack statistical robustness and tend to occur randomly, hence infrequently, see Figure \ref{fig:srbp}. Algorithm \ref{alg:SSIFN} illustrates this process, introducing an ensemble adjacency matrix $\it{\bf{A}}$, which amalgamates all adjacency matrices from bootstrapped sub-TMFGs.  After all repetitions, the occurrence probability of an edge between any pair is calculated, and only if this probability exceeds the defined confidence level (ConfLv) threshold, do we retain the edge. The final output $\mathbf{{S}}$ is a similarity matrix representing the original correlation, but with many structural zeros from the discarded edges to ensure sparsity.

\subsection{Bootstrapped Centrality Measures}

Subsequent experiments will utilize three centrality measures for portfolio weight calculation, including Degree Centrality, Communicability Betweenness Centrality, and Absolute Correlation. 

Degree Centrality is one of the simplest and most common centrality measures used to quantify the prominence of a node in a network. It is based on the idea that nodes with more connections (edges) to other nodes are more central and influential within the network. For an undirected network, the degree centrality, $c^d_i$ of a node i is calculated as the number of edges (connections) it has denoted by $k_i$. The normalized degree centrality is obtained by dividing $k_i$ by the maximum possible number of connections, which is $(n-1)$, where n is the total number of nodes in the network,
\begin{equation}
    c^d_i = k_i / (n - 1).
\end{equation}
This normalization allows for the comparison of centrality scores across different networks.

Communicability Betweenness Centrality (CBC) is an extension of the traditional betweenness centrality measure, which is based solely on the shortest paths between nodes. While betweenness centrality focuses on the number of shortest paths that pass through a given node, CBC takes into account the weighted sum of all paths between nodes, where the weight of each path is inversely proportional to its length. Mathematically, communicability between nodes i and j is calculated using the exponential of the adjacency matrix, $\mathbf{A}$, of the network. The adjacency matrix is a square matrix whose element $\mathbf{A}_{j,k}$ represents the connection between nodes i and j. The communicability between nodes i and j is given by the (j,k)-th element of the matrix exponential, denoted as $\exp{(\mathbf{A}_{j,k})}$. Communicability Betweenness Centrality is then calculated by summing the relative changes in communicability for all pairs of nodes when a node is removed from the network. For node i, the CBC is computed as:
\begin{equation}
    c^{CBC}_i = \frac{\sum_{i \neq j \neq k} {\exp{(\mathbf{A}_{j,k})}-\exp{(\mathbf{A}_{j,k}-E_i)}}}{\exp{(\mathbf{A}_{j,k})}}
\end{equation}
where $E_i$ is a matrix with the same dimensions as $\mathbf{A}_{j,k}$, representing the connections of node i(i.e., with 1s in the positions corresponding to the edges of node k and 0s elsewhere), and $(\mathbf{A}_{j,k}-E_i)$ represents a new adjacency matrix by removing node k from the network. In this formula, the summation is over all pairs of nodes i and j, excluding node k. The CBC quantifies the importance of node k by considering its role in facilitating communication between all pairs of nodes in the network, taking into account both direct and indirect paths.

The portfolio selection methods in the above section select assets with statistically significant decorrelation among the portfolio. Therefore an intuitive way for weights optimization is directly using the sum of absolute pairwise correlation, as the portfolio weight for each node/asset. Therefore, it is expressed as 
\begin{equation}
    c^{corr}_i = \sum_{j,i\neq j} {|C_{i,j}|},
\end{equation}
where $C_{i,j}$ represents the pairwise correlation between node i and node j.

A bootstrapping approach akin to Algorithm \ref{alg:SSIFN} is utilized for calculating statistically robust centrality. In each repetition, a sub-centrality, $c^t$, is determined within the sub-network obtained from bootstrapped observations, and the overall centrality is obtained by averaging all sub-centralities, as shown:
\begin{equation}
c_i = \frac{1}{t_r} \sum_{t=1}^{t_r}{c^t_i}
\end{equation}
where $c_i$ is the ensembled centrality, $c^t_i$ is the sub-centrality for node $i$, and $t_r$ is the number of repetitions.

\subsection{Portfolio Selection and Optimization} \label{sec:portSlcOpt}
In our application of the Statistically Robust Information Filtering Network (SR-IFN), we consider a total of $N$ assets with $T$ time-stamped historical observations. The resultant sparse similarity matrix, $\mathbf{{S}}$, represents the pairwise correlations between assets. Given the sparse nature of $\mathbf{{S}}$, it allows for the division of assets into two subsets: connected and disconnected. Disconnected assets lack any link to other assets, while connected ones possess at least one such link. By adjusting the confidence level (ConfLv) threshold within the SR-IFN, we can manipulate the quantities of disconnected and connected assets. For the purpose of establishing a portfolio with minimal correlation, we include all disconnected assets, while excluding the connected ones. More specifically, we select assets for which the sum of pairwise correlations is zero, as expressed in the following equation:
\begin{equation}\label{eq:slc}
\sum_{j, i\neq j} {{\mathbf{{S}}_{i,j}}} = 0.
\end{equation}
This results in the selection of assets that exhibit a very low statistical correlation with others within the portfolio.

To further enhance the portfolio, we optimize the weights such that they are inversely proportional to the ensembled centrality measures,
\begin{equation}\label{eq:opt}
w_i = \frac{1/c_i}{\sum_{j}{1/c_j}}
\end{equation}
where $w_i$ is the weight, and $c_i$ represents the centrality for asset $i$ in the portfolio. 

\begin{figure}[h!]
    \centering
    \includegraphics[width=0.4\textwidth]{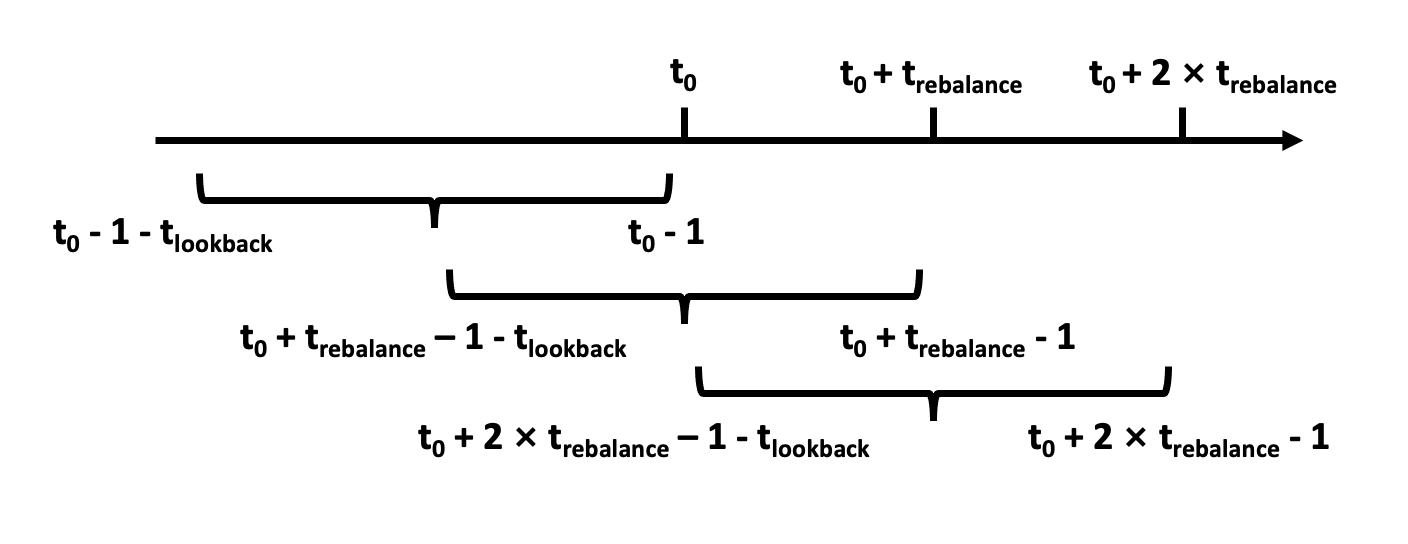}

    \caption{Portfolio is rebalanced every $t_{rebalance}$-day with a $t_{lookback}$ days look-back window of history.
}
\end{figure}\label{fig:rb}

Dynamic allocation of the portfolio is achieved through rebalancing every $t_{rebalance}$-day. The selection criteria in Equation \ref{eq:slc} and weights in Equation \ref{eq:opt} are re-calculated based on a $t_{lookback}$ days look-back window of history, see Figure \ref{fig:rb}.

\section{Implementation} \label{sec:imp}

\subsection{Data}
A series of experiments were conducted utilizing historical financial time-series data obtained from three principal capital markets: NASDAQ, FTSE, and HS300, spanning the period from January 1, 2010, to January 1, 2020. For each constituent stock, the daily log-return, denoted as $r_i(t)=\log(P_i(t))-\log(P_i(t-1))$, was computed using closing prices. Detailed statistics of the daily log-return distribution are furnished in Table \ref{tab:sts} for subsequent comparison and discussion.

\begin{table}[h!]
\center
\begin{tabular}{c|c|c|c}
\toprule
& {NASDAQ} & {FTSE} & {HS300} \\
\midrule
Ann. Return & 16.0\% & 5.8\% & 9.9\% \\
       \midrule
Ann. Std.Dev. &  17.5\% & 15.8\% & 23.2\% \\
       \midrule
D. Skewness  &  -0.44 & -0.95 & -0.89 \\
       \midrule
Max. Drawdown & -24.0\% & -43.5\% & -52.3\%\\
\bottomrule
\end{tabular}
\vspace{0.5pt}
    \caption{Statistics table for the log return distribution in NASDAQ, FTSE and HS300 between 01/01/2010 and 01/01/2020, including annualized return mean, annualized return standard deviation, daily return skewness, and maximum drawdown.}\label{tab:sts}
\end{table}

The chosen indexes are emblematic of distinctly divergent market dynamics during the designated period. NASDAQ was in a phase colloquially referred to as its 'golden ten years', characterized by a substantial annualized mean return and moderate volatility. The skewness of the return distribution is less negative compared to the other indexes, indicating fewer extreme loss events and consequently, a lower maximum drawdown throughout this period. In contrast, both FTSE and HS300 exhibited a high negative skewness and substantial drawdown over the same period. Additionally, the FTSE was more conservative, with a lower average return and volatility, whereas the HS300 displayed considerably higher volatility.

\subsection{Experiment Setup}\label{sec:expsetup}

This section is devoted to the selection of portfolios exclusively from an index component stock pool. Consequently, the weights of the portfolio are maintained at $1/N$, where $N$ represents the total number of assets in the chosen portfolio, for the sake of simplicity and controlled comparison. The portfolio undergoes rebalancing every $t_{\text{rebalance}}$ days, with a historical look-back period of $t_{\text{lookback}}$ days for the measurement of empirical correlation and other historical statistical properties. Experiments are included both with and without transaction costs of 20 basis points (bps) to simulate commission and bid-ask spread costs. The complete period is partitioned into in-sample and out-of-sample periods before and after 01/01/2017. A grid search over $t_{\text{rebalance}}$ and $t_{\text{lookback}}$ is conducted in-sample for analysis and optimization, and the optimal parameters are retained for the out-of-sample period to showcase the persistence and significance of the method. In addition, to demonstrate robustness and present statistics, all experiments across the three markets are repeated and ensembled with varying starting dates.

\begin{figure}[h!]
    \centering
    \includegraphics[width=0.45\textwidth]{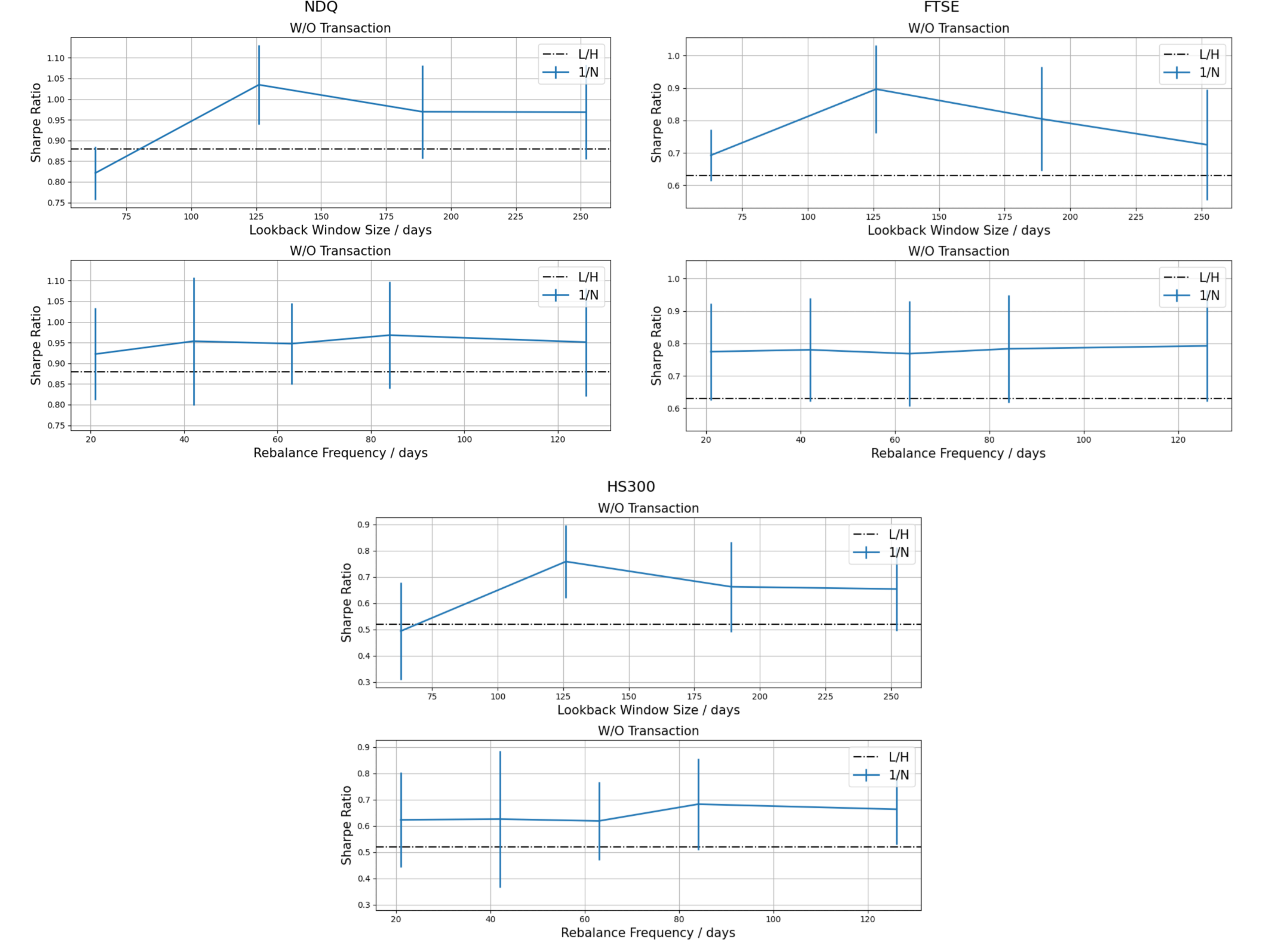}

    \caption{Grid search results in Sharpe Ratio across NASDAQ (top left), FTSE (top right), and HS300 (bottom) over various rebalance frequencies and lookback window sizes. The Left and right columns show results with and without 20bps transaction costs. L/H represents the long-hold portfolio over the entire stock pool, and 1/N represents the simple equally weighted selected portfolio averaged over other parameters. The grid search results of Confidence Levels of SR-IFN are not shown as it is not optimized for the remaining experiments.
}
\end{figure}\label{fig:gc}

Figure \ref{fig:gc} portrays the grid search results over a range of rebalance frequencies, $t_{\text{rebalance}}$, and lookback window sizes, $t_{\text{lookback}}$. The results are averaged across varying $t_{\text{rebalance}}$, $t_{\text{lookback}}$ and Confidence Levels of Statistically Robust Information Filtering Network (SR-IFN, denoted as ConfLv), but the outcomes in relation to ConfLv are not displayed as it is not optimized for the remaining experiments. This grid search is executed in-sample from 01/01/2010 to 01/01/2017. While it is safe to assume that $t_{\text{lookback}}$ is optimally at 126 days for all three markets, there is a minor discrepancy among $t_{\text{rebalance}}$. Nonetheless, for simplicity and consistency, we employ an 84-day $t_{\text{rebalance}}$ for the remaining experiments.

\section{Results}
\subsection{Topological Portfolio Selection} \label{sec:portSelectResults}

The Statistically Robust Information Filtering Network (SR-IFN), introduced in Section~\ref{sec:SSifn}, provides a statistically robust selection predicated on historical correlation. The remaining correlated features, derived from the historical period, due to their robustness, are anticipated to maintain their correlation for a brief future period. This intrinsic ability to predict future correlation serves as a pivotal criterion for numerous portfolio selection and optimization techniques, as their primary objective is to identify the least correlated portfolio. In this section, we scrutinize the influence of the peripheral portfolio over the central portfolio, as defined by the correlation graph, and exhibit the supplemental gain from the SR-IFN peripheral portfolio in comparison to a classic correlation-based portfolio.

\begin{figure*}[h!]
    \centering
    \includegraphics[width=0.7\textwidth]{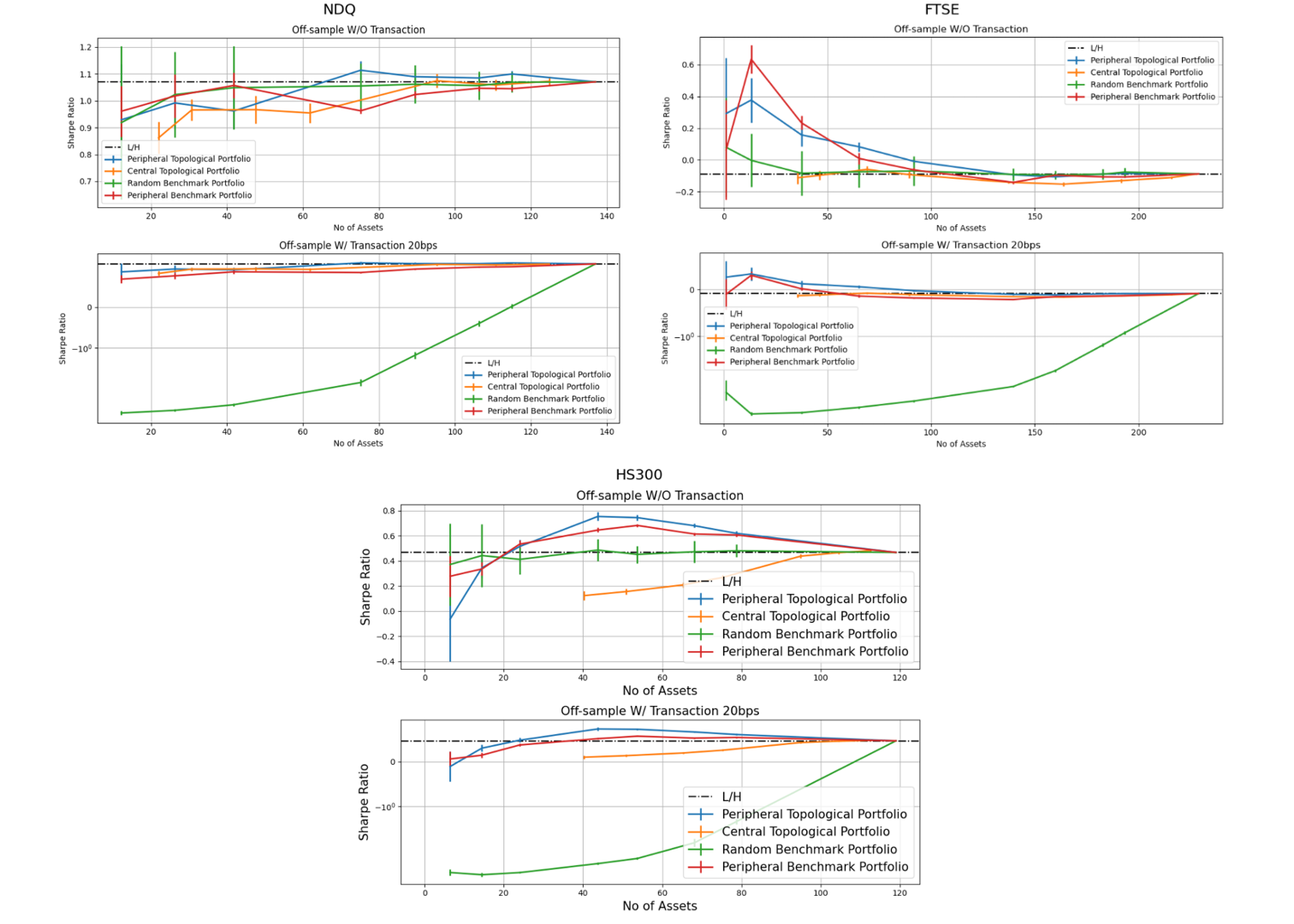}

    \caption{Portfolio selection results in Sharpe Ratio across NASDAQ (top left), FTSE (top right), and HS300 (bottom). The selection is performed based on the SR-IFN, and we report the central (orange) and peripheral (blue) portfolios as well as a random subsampled (green) portfolio to showcase the efficacy of the peripheral one. For each subplot, the top and bottom (with symmetrical log scale) rows represent without and with 20bps transaction fees for the out-of-sample period(2017-2020). The rebalance window is fixed at 84 days and the lookback window is fixed at 126 days, optimized in the in-sample period. The peripheral and central portfolios do not have the exact same number of assets in comparison, as SR-IFN selects based on confidence level instead of an exact parameter.
}\label{fig:portSelect}
\end{figure*}

Figure \ref{fig:portSelect} showcases the results in terms of the Sharpe Ratio for NASDAQ, FTSE, and HS300. The parameters $t_{\text{rebalance}}=84$ and $t_{\text{lookback}}=126$ are fixed, which are optimized in-sample from section \ref{sec:expsetup}, and the outcomes of the out-of-sample period are presented. For each subplot, the top row and bottom row represent experiments without and with 20bps transaction costs, with the bottom row being displayed in a symmetrical log scale for enhanced visualization and comparison. The Peripheral Topological Portfolio (PTP) in blue is the principal portfolio selected by SR-IFN. By varying the ConfLv of SR-IFN, we illustrate the performance with respect to different numbers of assets. Given that PTP is constructed from the disconnected assets of the correlation graph, its counterpart, Central Topological Portfolio (CTP) in yellow, represented by the connected assets from the correlation graph, is also exhibited as a supplement to portray the nearly symmetrical gain and loss. Since the number of assets is not a direct parameter in the algorithm but is controlled by the ConfLv, PTP and CTP do not have the exact same number of assets when comparing the two curves. Moreover, to showcase the efficacy of PTP, we also present the results for a randomly sub-sampled portfolio with the same number of assets as PTP, denoted as Random Benchmark Portfolio (RBP) in green, a Peripheral Benchmark Portfolio (PBP) in red that is constructed by selecting the assets with the least sum of pairwise correlation, as well as a simple long hold strategy represented by a dashed line.

In all three markets, the two peripheral portfolios, PTP and PBP, both yield superior performance compared to RBP and CTP, suggesting a clear advantage in adopting the peripheral portfolio as discussed in section \ref{sec:lt_pso}. For out-of-sample experiments, PTP surpasses PBP when the number of assets is relatively large with no transaction cost, and if a 20bps transaction cost is applied, the range where PTP outperforms PBP extends. These findings corroborate that PTP is superior to the benchmark PBP with statistical significance and consistency across markets and conditions. Specifically, SR-IFN provides a more robust mechanism to identify assets with the most/least correlation than the simple empirical correlation method, and this effect is more likely to persist in the future. Furthermore, by contrasting PTP and CTP, the gains and losses are roughly symmetrical around the Long/Hold dashed line, suggesting that the gain in PTP predominantly arises from selecting the peripheral assets as opposed to other factors.

To further refine our discussion, we place a restriction on our portfolio size to include more than 50 and less than 100 assets, aiming to mitigate the high variance at the tail of the performance distribution. When examining the out-of-sample period without transaction fees, the average Sharpe Ratio for NASDAQ is 1.10, while with the inclusion of 20bps transaction fees, it marginally decreases to 1.08. This is compared against a Long/Hold (L/H) benchmark of 1.07. In a similar vein, the FTSE index records an average Sharpe Ratio of 0.28 without transaction fees and 0.24 with these fees, against an L/H benchmark of -0.09. The HS300 index exhibits a Sharpe Ratio of 0.65 without transaction fees and 0.62 with the inclusion of 20bps transaction fees, compared to an L/H benchmark of 0.47. Thus, the net gain in the out-of-sample Sharpe Ratio equates to approximately 3\% and 1\% in the case of NASDAQ, 411\%, and 367\% for FTSE, and 38\% and 32\% for HS300, without and with transaction fees respectively.

\begin{figure*}[h!]
    \centering
    \includegraphics[width=0.7\textwidth]{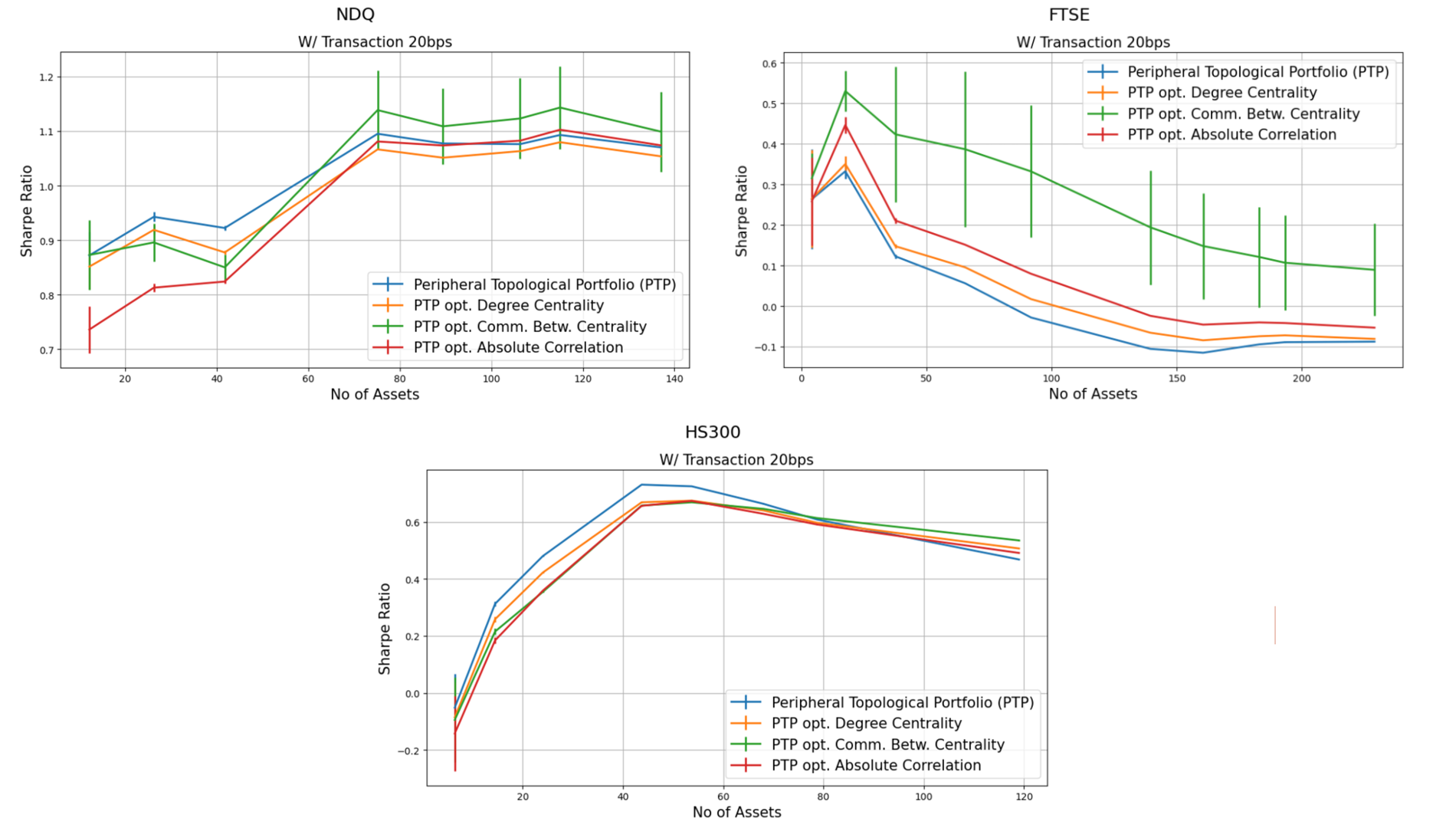}

    \caption{Portfolio optimization results for the out-of-sample period (2017-2020) in Sharpe Ratio across NASDAQ (top left), FTSE (top right), and HS300 (bottom) with 20bps transaction fee. We report the original Peripheral Topological Portfolio of equal weights as the benchmark (blue), and the optimized PTPs whose weights are inversely proportional to Degree Centrality (yellow), Communicability Betweenness Centrality (green), and Absolute Correlation (red). The rebalance window is fixed at 84 days and the lookback window is fixed at 126 days, optimized in the in-sample period. 
}\label{fig:portOpt}
\end{figure*}

The noteworthy performance observed in the FTSE and HS300 indices can likely be attributed to the specific market dynamics during the chosen period, characterized by a highly negative skew and a substantial maximum drawdown. For instance, during the 'golden period' of NASDAQ, a high beta and generally high correlation among the index component stocks result in a low signal-to-noise ratio when identifying the least correlated stocks. In contrast, in the more turbulent and less bullish market dynamics observed in the FTSE and HS300 indices, where the correlation among component stocks presents greater diversity, the underlying SR-IFN results in more significant findings in terms of correlation filtering and inference. In essence, SR-IFN for portfolio selection generally has a positive impact on the selection of the least correlated assets, leading to improved portfolio performance. This performance is more pronounced when the underlying market dynamic is less bullish and subject to more extreme losses.

\subsection{Topological Portfolio Optimization}\label{sec:portOptResults}

\begin{table*}[h!]
\center
\setlength\tabcolsep{2pt}
\begin{tabular}{c|ccc|ccc|ccc|}
\toprule
& \multicolumn{3}{|c|}{NASDAQ} & \multicolumn{3}{|c|}{FTSE} & \multicolumn{3}{|c|}{HS300} \\
\midrule

& L/H & PTP & PTP+CBC & L/H & PTP & PTP+CBC & L/H & PTP & PTP+CBC  \\
\midrule
Ann. Return & 16.6\% & 16.2\% & 15.7\% &
            -1.6\% & 1.1\% & 5.5\% &
            8.7\% & 11.5\% & 12.1\% 
            
\\
       \midrule
Ann. Std.Dev. &  15.5\% & 15.0\% & 14.1\% &
                18.4\% & 19.5\% & 17.7\% &
                18.6\% & 18.7\% & 18.8\% 
\\
\midrule
Sharpe R. & 1.07   & 1.08 & 1.12&
                -0.09 &  0.24&  0.42&
                0.47 &  0.62&  0.65

\\
       \midrule
Max. Drawdown & -23.2\% & -22.1\% & -18.1\%&
                 -43.5\% & -48.5\% & -42.4\%&
                  -30.0\% & -28.0\% & -29.0\%

\\
\bottomrule
\end{tabular}
\vspace{0.5pt}
    \caption{Aggregated performance statistics of L/H benchmark, Peripheral Topological Portfolio (PTP) and PTP optimised by Communicability Betweenness Centrality (CBC) in NASDAQ, FTSE and HS300. The table reports averaged statistics for portfolios with a number of assets between 50 and 100, including the annualized mean return, annualized return standard deviation, annualized Sharpe Ratio, daily return skewness and maximum drawdown.}\label{tab:port}
\end{table*}

In the preceding section, we have demonstrated the significant advantage of a peripheral portfolio selection strategy in the context of high drawdown periods. This strategy, founded on the least correlated portfolio, can be further refined by incorporating other topological properties to optimize the weighting of the selected portfolio. Herein, we continue to underscore the merit of a more peripheral portfolio, characterized by reduced correlation and superior performance, by assigning weights that are inversely proportional to centrality measures within the previously selected Peripheral Topological Portfolio (PTP). Degree Centrality, as one of the simplest measures of centrality, and Communicability Betweenness Centrality, a more complex but well-documented measure, are included in our study. Furthermore, given its intuitive nature and alignment with the overall theme of decorrelation, Absolute Correlation is also incorporated into our experiments.

Figure \ref{fig:portOpt} depicts the performance across the three markets in terms of the Sharpe Ratio. Within each market, we plot PTP, serving as the benchmark, and three optimized versions of PTP wherein weights are inversely proportional to centrality measures, including Degree Centrality (yellow), Communicability Betweenness Centrality (green), and Absolute Correlation (red).

Maintaining the same comparative framework, we restrict our analysis to portfolios comprising more than 50 and less than 100 assets, in order to mitigate the high variance at the tail of the performance distribution. For brevity, we limit our analysis to experiments incorporating 20 bps transaction fees. For the NASDAQ index, the average Sharpe Ratio is improved from 1.08 to 1.12, representing an approximate 4\% improvement when optimized by Communicability Betweenness Centrality. Absolute Correlation yields an equal 1.08, while Degree Centrality results in a slightly inferior 1.06. For the FTSE index, Communicability Betweenness Centrality optimization improves the Sharpe Ratio from 0.24 to 0.42, an impressive 75\% enhancement, while Absolute Correlation and Degree Centrality yield improvements to 0.31 (29\%) and 0.25 (4\%), respectively. For the HS300 index, the average Sharpe Ratio improves from 0.62 to 0.65 (approximately 5\%) when optimized by Communicability Betweenness Centrality, while remaining unchanged under the other two methods.

Furthermore in Table \ref{tab:port}, we summarize the performance of selection in the previous subsection and optimization in this subsection by reporting the aggregated performance for the Long/Hold benchmark, PTP and PTP optimized by CBC (PTP+CBC). Apart from a superior Sharpe Ratio discussed above, our PTP+CBC demonstrates significant improvement in the risk matrices. PTP+CBC has reduced annualized return standard deviation by 1.4\% in NASDAQ, 0.7\% in FTSE and kept similar in HS300, as well as shrank the maximum drawdown by 5.1\% in NASDAQ, 1.1\% in FTSE and 1\% in HS300.

This section illustrates the impact of weighting the portfolio inversely proportional to different centrality measures. We provide quantitative evidence of a robust improvement over the simple, equally-weighted PTP.  Furthermore, Figure \ref{fig:portOpt} demonstrates that, apart from the FTSE index where the effect is consistently dominant across all asset numbers, the effect is particularly pronounced for portfolios with larger asset numbers. One plausible explanation is that, for smaller PTPs, the assets are already optimally selected and much of the topological information has been extracted. As a result, additional optimization may suffer from a low signal-to-noise ratio, as the application of infinitesimally small weights effectively equates to deselection. This hypothesis aligns with the more pronounced effect observed in the FTSE index, given its effectively larger pool of component stocks compared to the other two indices.

\section{Conclusion}

In this study, we present a novel, statistically robust bootstrapping method designed to select a robust structure from bootstrapped information filtering networks, hereby referred to as the Statistically Robust Information Filtering Network (SR-IFN). This method improves upon the existing Information Filtering Network (IFN) by reducing redundant edges formed due to applied graphical constraints. The SR-IFN accepts multivariate observations as inputs and outputs a sparse similarity matrix and a network, both of which are subsequently employed for portfolio selection and optimization with constant rebalancing. Our experiments spanned a decade-long history across three distinct markets, utilizing the first 70\% of the data to select parameters such as rebalancing frequency and lookback window size. The results reported are based on the remaining three years of out-of-sample data.  Our in-sample grid search for parameter tuning demonstrated consistent outperformance of the benchmark, mirroring the findings in the out-of-sample period, thereby reinforcing the robustness of the proposed method in even the most challenging financial applications.

Our findings indicate that the deployment of such an innovative approach results in a Sharpe Ratio improvement of 1\%, 367\%, and 32\% with 20bps transaction costs for market indices. This is achieved by simply selecting a subset of composite stocks in the US, UK, and China markets, respectively. Moreover, the performance can be further amplified by optimizing the portfolio weights based on the centrality measures of the output network, yielding additional improvements of 4\%, 75\%, and 5\%. The cumulative improvement derived from both approaches enhances the results by 5\%, 567\%, and 38\% for the NASDAQ, FTSE, and HS300 indices, respectively. The disparities in the magnitude of improvement are likely attributed to the market dynamics of the selected period. For instance, NASDAQ was in its 'golden period', while the other two markets underwent significant drawdowns. Consequently, further improvement of an already efficient system (NASDAQ) proved more challenging than the other two, which serves as a testament to the method's resilience under extreme market conditions. Furthermore, despite a marginal boost in the risk-adjusted reward in NASDAQ compared to the other two markets, the risk metrics are notably reduced in both annualized standard deviation of 1.4\% and maximum drawdown of 5.1\%. Additional findings reveal that the underlying method performs well with large-dimension data (number of assets) with computational efficiency.

\newpage
  
\newpage

\bibliographystyle{ACM-Reference-Format}
\bibliography{reference}

\end{document}